\documentclass[prd, twocolumn, superscriptaddress, nofootinbib]{revtex4}
\usepackage{epsfig}

\newcommand{\beq}{\begin{equation}}
\newcommand{\eeq}{\end{equation}}
\newcommand{\beqa}{\begin{eqnarray}}
\newcommand{\eeqa}{\end{eqnarray}}
\newcommand{\om}{\Omega_m}

\newcommand{\alp}{Alcock-Paczy{\'n}ski\ } 
\newcommand{\aandp}{Alcock \& Paczy{\'n}ski\ } 

\def\la{\mathrel{\mathpalette\fun <}}

\def\fun#1#2{\lower3.6pt\vbox{\baselineskip0pt\lineskip.9pt
  \ialign{$\mathsurround=0pt#1\hfil##\hfil$\crcr#2\crcr\sim\crcr}}}

\begin{document} 

\title{Cosmic Shear with Next Generation Redshift Surveys 
as a Cosmological Probe} 
\author{Eric V. Linder}
\affiliation{Physics Division, Lawrence Berkeley National Laboratory, 
Berkeley, CA 94720} 

\begin{abstract} 
The expansion of the universe causes spacetime curvature, 
distinguishing between distances measured along and transverse 
to the line of sight. The ratio of these distances, e.g.~the cosmic 
shear distortion of a sphere defined by observations of large scale 
structure as suggested by Alcock \& Paczy{\'n}ski, provides a method 
for exploring the expansion as a function of redshift.  The theoretical 
sensitivity to cosmological parameters, including the dark energy 
equation of state, is presented. Remarkably, 
sensitivity to the time variation of the dark energy equation of 
state is best achieved by observations at redshifts $z\la 1$.  While 
systematic errors greatly degrade the theoretical sensitivity, 
this probe may still offer useful parameter estimation, especially 
in complementarity with a distance measure like the Type Ia 
supernova method implemented by SNAP.  Possible future observations 
of the Alcock-Paczy{\'n}ski distortion by the KAOS project on a 8 
meter ground based telescope are considered. 
\end{abstract} 

\maketitle 

\section{Introduction} \label{sec.intro}

We now have strong evidence that the expansion of the universe is accelerating, 
from the original method of Type Ia supernova distance-redshift measurements 
\cite{perl99,riess98} and concordant observations of the cosmic microwave 
background (CMB) power spectrum and of large scale structure \cite{bond,perc}.  
Understanding the 
nature of the dark energy responsible for the acceleration will have profound 
implications for cosmology, high energy physics, and fundamental physics.  
Mapping 
the expansion history of the universe offers a way to gain insights into the 
dark energy and the fate of the universe, for example by characterizing the 
equation of state behavior which is directly related to properties of the 
scalar field potential. 

Distance measures, notably the supernova method, have proved useful at 
constraining 
the energy density and equation of state of the dark energy, with great 
improvements 
expected in the next decade.  But these involve an integration over the 
expansion 
rate behavior $H(z)$, which itself involves a redshift integral over the 
equation 
of state $w(z)$.  We can ask whether we can devise a more direct probe of the 
acceleration.  In fact one such, the redshift drift test, was proposed by 
Sandage 
\cite{san61} in 1961 and developed further by Linder \cite{lin97}. 

The redshift of a source is a central astrophysical observable.  It is 
directly related to 
the change in time intervals due to the cosmic expansion between a photon's 
emission and observation, $z=dt_o/dt_e-1=a_o/a_e-1$, 
where $a(t)$ is the scale factor of the universe.  But obviously one could 
consider a second derivative term, a time dependence of the redshift 
itself as the universe ages: 
\beqa 
{dz\over dt_o}&=&{d\over dt_o}\left[{a(t_o)\over a(t_e)}\right]=[\dot a(t_o)- 
\dot a(t_e)]/a(t_e) \\ 
&=&H_0(1+z)-H(z). 
\label{dzt}\eeqa 

This provides a direct measure of acceleration, being effectively a 
second time derivative, as can be seen from the low redshift 
limit: $\Delta z/z\approx -q_0H_0\Delta t$, 
where $q_0$ is the present deceleration parameter. 
However since the astronomical observing time is much smaller than the 
Hubble time, $\Delta t\ll H_0^{-1}$, 
this is not a practical probe.  For example in the most optimistic case 
of observing over a period of 10 years a hypothetical spectral line 
emitted at the CMB last scattering 
surface at $z=10^3$, one requires a redshift measurement of a part 
in $10^5$ to distinguish cosmological models.  For emission line objects at 
$z=5$, this becomes a part in $10^8$.  

Moreover, just as peculiar velocities affect 
redshift measurements at the level of $v/c$, so do peculiar accelerations, 
i.e.~local gravitational potentials $\Phi$ from inhomogeneously distributed 
matter, 
affect the redshift drift measurements at a level $\Phi/c^2\approx10^{-5}$. 
This latter effect even ruins the generalization of the redshift drift called 
the cosmic pulsar test, where timing is improved by measuring a large number 
$N$ of wavelengths or pulses \cite{lin97}. 

But if we are stymied in measuring the acceleration directly, at least we can 
hope to measure the first derivative of the expansion, $H(z)$.  One of the 
ways to do this is the cosmic shear, or {\alp} test.  Section \ref{sec.form} 
sets up the formalism while Section \ref{sec.sens} applies Fisher matrix 
analysis to investigate the theoretical sensitivity of this method for 
estimating the 
cosmological parameters.  In Section \ref{sec.sys} we introduce observational 
sanity in the form of systematic uncertainties and discuss the proposed KAOS 
project as a means of carrying out this test.  We summarize our conclusions 
and plans for future work in Section \ref{sec.concl}. 

\section{Cosmic Shear Test} \label{sec.form}

Proper distances measured along the line of sight carry information through 
light 
emitted at different times in the expansion history of the universe.  Therefore 
in 
a sphere of comoving points, differences in the emission times lead to probing 
the geometry at different expansion rates.  So observationally a sphere will 
appear 
to be distorted, or sheared, with the magnitude of the effect sensitive to the 
expansion rate.  This is the cosmic shear effect discussed by \aandp 
(\cite{ap}; do not confuse this with shear from weak gravitational 
lensing -- due to inhomogeneities rather than the global structure of 
spacetime). 

In more detail, if we consider the 
small difference in radial distance between nearby emitters, then we localize 
the 
behavior in redshift and essentially measure the expansion rate at that time. 
However 
transverse to the radial direction the angular separation between comoving 
points is 
measured at a constant value of the scale factor $a(t_e)$, and then 
the light from each 
source propagates over the same intervening distance to the observer 
at $t_o$.  Thus in the 
radial case the data gives a snapshot of the expansion rate while the transverse 
distance contains an integration over the expansion history from emission to 
observation. 

Consider a sphere of comoving points.  The distance through the center of the 
sphere 
along the line of sight is the proper distance, 
\beq
dr_\parallel=dt=a(t)\,dr_c=(1+z)^{-1}H^{-1}(z)\,\Delta z.
\eeq
The transverse distance is 
\beq
dr_\perp=r_a(z)\,\Delta\theta=(1+z)^{-1}r_c(z)\,\Delta\theta,
\eeq
where $z=a^{-1}-1$ is the redshift, $H=\dot a/a$ is the Hubble parameter, $r_c$ 
the 
comoving distance, and $r_a$ the angular distance (see, for example, 
\cite{lin97}).  

From the observables of the angular scale $\Delta\theta$ of such comoving 
sources, 
their central redshift $z$, and their redshift extent $\Delta z$, one can form a 
quantity 
\beq
D(z)\equiv\Delta z/\Delta\theta=H(z)\int_0^z dz'/H(z').
\label{heta}
\eeq 
The cosmic shear is then 
\beq 
S=\sqrt{2\left[1-\left({\Delta z\over \Delta\theta}\right)^2\right]}.
\label{shear}\eeq 

This has some excellent properties for a cosmological probe. In particular, it 
has 
dependence on $H(z)$ directly, not just through an integral.  Since $H$ is 
related to the energy density of the universe 
then one can try to map the density history and equation of state. 
Indeed for time varying equation of state of the dark energy $w(z)$, 
$H$ depends on an integral of $w(z)$, so distance measures involve 
$w(z)$ as a double integral.  Therefore one might hope that 
the \alp differential distance test might be more sensitive to 
reconstructing the dark energy equation of state than a standard 
distance test. 

Another interesting characteristic of (\ref{heta}) is that it does 
not depend on any absolute scale, since $H$ appears in both numerator 
and denominator. So there is no absolute measurement to marginalize over. 

The physics of the test seems clean, using pure geometry of the background 
spacetime, 
so long as we can find sources that are comoving and defining a known local 
spatial 
geometry. Conventionally the local source geometry is taken to be spherical, as 
it would be 
for an isotropic arrangement. Possibilities for defining isotropically arranged 
sources 
include large coherent objects such as superclusters in the linear density 
regime and 
so following the isotropic expansion of the universe, or lengths defined through 
correlation functions between individual objects, such as for galaxies or  
Lyman 
alpha forest absorbers. The extent to which these assumptions break down or 
cannot 
be corrected yield systematic errors to plague the method.  As for many other  
cosmological probes, the systematic errors turn out to be more severe than 
statistical 
errors from insufficient data. Section \ref{sec.sys} discusses this further. 

\section{Sensitivity to Cosmological Parameters} \label{sec.sens} 

We begin by considering a purely theoretical analysis of the capabilities of the 
test, 
leaving observational realities for \S\ref{sec.sys}. A good way to understand 
the 
sensitivities and degeneracies of cosmological probes is through Fisher matrix 
analysis of the dependence of the observable on the parameters. We take a flat 
universe defined by three parameters: the dimensionless matter density $\om$ (so 
the 
dark energy density is $1-\om$), the value of the dark energy equation of state 
today 
$w_0$, and a measure of its time variation $w'$. Since we want to consider 
observations extending to redshifts greater than unity, e.g.~Lyman alpha 
observations are most plentiful with $z\approx3$, we adopt the equation 
of state parametrization $w(z)=w_0+w_az/(1+z)$ with the definition 
$w'=dw/d\ln(1+z)|_{z=1}=w_a/2$ 
\cite{linwa}. This approximates well the behavior of several classes of 
dark energy models, 
especially those with a slow roll phase, is well behaved even for 
$z>1$, and allows insight 
into the effects of the physically expected time variation in the equation of 
state. 

The distance distortion $D(z)= H(z)\int dz/H$ differs in its behavior in an 
interesting way from distance or volume measures: the two factors actually 
depend 
on the cosmological parameters in opposite ways because they have reciprocal 
dependence on the expansion rate. At low redshift the direct $H(z)$ dependence 
dominates since all distances must be similar. But at high redshifts the 
universe was 
matter dominated and so $H(z\gg1)$ is insensitive to the equation of state 
parameters and the distance factor takes over, since it retains memory of those 
parameters due to its integral nature. 

\begin{figure}[!hbt]
\begin{center} 
\psfig{file=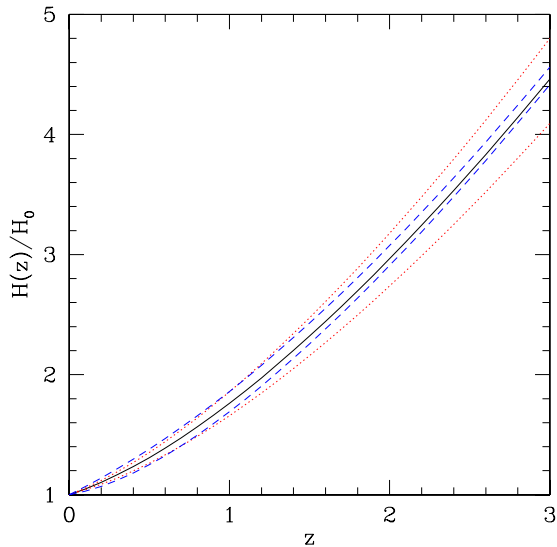,width=3in} 
\caption{The Hubble parameter $H(z)$ as a function of redshift, 
for the fiducial flat model $\om=0.3$, $w=-1$ (solid curve) and 
variants.  Upper (lower) dotted curves have $\om=0.35$ (0.25); 
upper (lower) dashed curves have $w=-0.8$ ($-1.2$). 
} 
\label{fig.hz}
\end{center} 
\end{figure}

\begin{figure}[!hbt]
\begin{center} 
\psfig{file=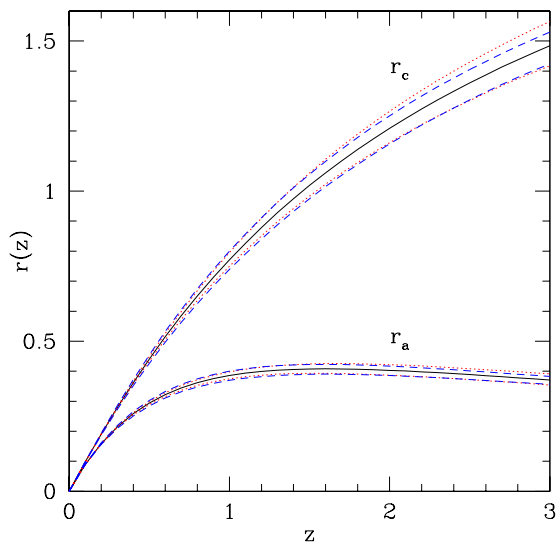,width=3in} 
\caption{The comoving distance $r_c$ and angular 
distance $r_a$ as a function of redshift, for the same models 
as Fig.~\ref{fig.hz}.  However now upper and lower models are 
interchanged, e.g.~the {\it upper} dotted curve has $\om=0.25$ etc.
} 
\label{fig.rz}
\end{center} 
\end{figure}

Figures \ref{fig.hz} and \ref{fig.rz} show the dependence of the Hubble 
parameter and 
the distances $r_c$ and $r_a$, respectively, on the cosmological 
parameters 
$\om$ and $w_0$ ($w_a$ behaves similarly). One clearly sees that 
the dependencies are in inverse relation. This 
implies that there can be crossover ranges in redshift where the distance 
distortion 
$D$ is essentially independent of one of the parameters. While this makes it 
impossible to determine that aspect of cosmology with observations in that 
redshift range, it has a benefit as well. From the figures one can 
deduce that degeneracies exist 
where one parameter can be adjusted to counteract the effect of another. 
But at the 
crossover points one parameter will {\it not} affect the distortion and so the 
degeneracy can be broken.  Essentially, observations near a crossover apply to a 
reduced phase space, and hence the other parameter estimates will be sharper. 
This 
is clearly seen in later figures. 

\begin{figure}[!hbt]
\begin{center} 
\psfig{file=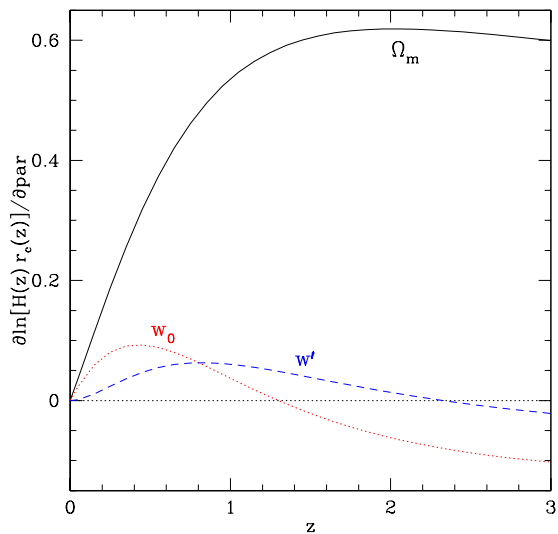,width=3in} 
\caption{The sensitivity of the observable, here 
$D=H(z)\,r_c(z)$, to the cosmological parameters is encoded in the 
derivatives plotted here.  The larger the absolute magnitude of the 
derivative at a particular redshift, the more constraining the 
observations there. 
} 
\label{fig.sens}
\end{center} 
\end{figure}

An excellent way to explore the sensitivity of an observational test to the 
cosmology is through Fisher matrix analysis \cite{tegfis}.  This 
methodology approximates the likelihood surface of the parameter fit 
to observations with a gaussian probability near the best fit, the fiducial 
model. The 
sensitivity of its estimation of parameter values depends on the derivatives of 
the 
observational quantity with respect to the parameter, 
$\partial D/\partial x$, and the 
precision with which the observations can be made. Writing the errors as $\delta 
D=(\delta D/D)D$, we see that for a given fractional measurement precision 
$\sigma_D\equiv\delta D/D$ we can obtain a parameter error estimate $\delta x$ 
by investigating 
$\partial\ln D/\partial x$. Figure \ref{fig.sens} shows this central quantity of 
Fisher analysis. 

Indeed we can read off almost all of our results from this 
sensitivity graph.  The previously discussed crossovers are clear, 
with the probe aloof to $w_0$ around $z\approx1.3$ and to $w'$ 
(i.e.~$w_a$) around $z\approx2.3$. Sensitivity to $\om$ grows out 
to $z\approx1$, after which it levels off; $w_0$ has a sweet spot 
at $z\approx0.4$ 
and again above $z\approx2$; the time variation $w'$ has a broad, though low, 
impact from $z\approx0.6-1.1$. However at $z\approx1.3$ we expect $w'$ to be 
uncorrelated with $w_0$ and so possibly easier to determine. 

Also note that while at low redshift the sensitivities to $\om$ and $w_0$ enter 
with 
the same sign, this changes after the $w_0$ crossover.  This means that at low 
redshift the error ellipse in the $w_0-\om$ plane will have the same orientation 
as for 
the supernova distance case: making $w_0$ larger (less negative) can be 
counteracted by making $\om$ smaller, defining a degeneracy direction in that 
plane. However higher redshift cosmic shear observations will have an orthogonal 
degeneracy direction, holding out the promise of complementarity with 
supernovae. 
A similar rotation of error ellipse contours with redshift can be predicted 
between 
$w_0$ and $w'$. 

Figure \ref{fig.sens} for the Fisher sensitivity even allows us to calculate 
lower limits on 
the precision with which we can estimate the parameters:
\beq
\delta x>(\partial\ln D/\partial x)^{-1}\sigma_D.
\eeq
This lower limit refers to fixing all parameters but one, and so will 
underestimate the 
true error due to degeneracies. If we take, say, 1\% precision in observations, 
then 
the lower limits on estimating $\{\om,w_0,w'\}$ are 0.015, 0.1, 0.15 for the 
peaks 
of the sensitivity curves.  However multiple observations can improve on the 
precision, while systematic errors will put a floor on it. To proceed further 
quantitatively we must input an observational suite into the Fisher method. 

To best illustrate the results we begin with a simple set of observations. We 
assume equal numbers $n$ of observations in redshift bins of width 0.1, 
each with the same precision $\sigma_D$, and vary the redshift range the 
observations cover. Initially we apply only statistical errors and so only the 
combination $\sigma_D/\sqrt{n}$ matters. 

\begin{figure}[!hbt]
\begin{center} 
\psfig{file=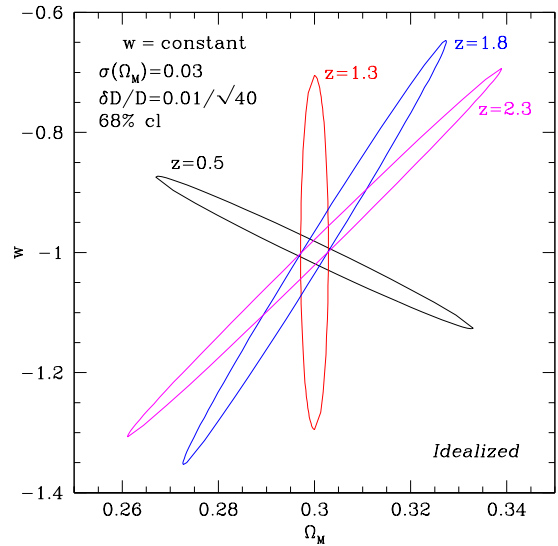,width=3in} 
\caption{Idealized results for estimation of $w$ and $\om$, with only 
statistical errors.  Note the equation 
of state is fixed to be constant a priori.  This is purely illustrative, showing 
the rotation of degeneracy directions and decorrelation at the crossover 
redshift. 
The sizes of the ellipses are idealized, corresponding 
to pure statistics for 40 observations of precision 1\% in a 0.1 bin in 
redshift. 
} 
\label{fig.idealomw}
\end{center} 
\end{figure}

\begin{figure}[!hbt]
\begin{center} 
\psfig{file=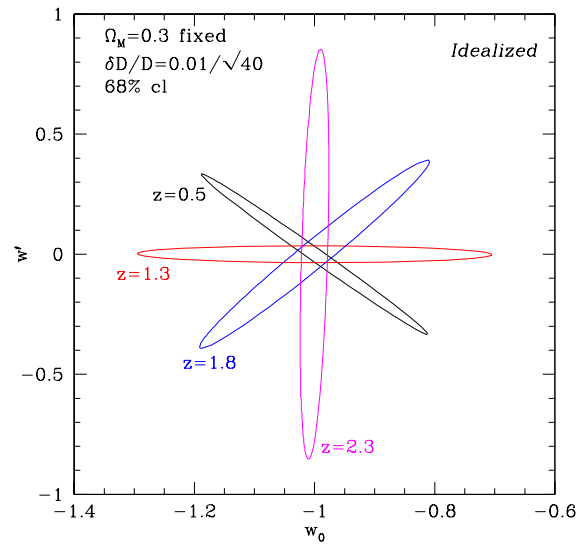,width=3in} 
\caption{Idealized results for estimating $w'$ and $w_0$, with $\om$ fixed 
and only statistical errors. This is purely 
illustrative, showing the rotation of degeneracy directions and decorrelation 
at the crossover redshifts. The sizes of the ellipses are idealized, 
corresponding 
to pure statistics for 40 observations of precision 1\% in a 0.1 bin in 
redshift. 
} 
\label{fig.idealwwp}
\end{center} 
\end{figure}

Purely to test our intuition from the Fisher sensitivity figure, we take the 
highly 
idealized situation of constant $w(z)$ (i.e.~fix $w'=0$ {\it a priori}) in 
Figure 
\ref{fig.idealomw} and fixed $\om$ in Figure \ref{fig.idealwwp}, as well as 
unrealistically good precision. The rotation of the contours with 
redshift is clear. 
Note that for the crossover redshift $z\approx1.3$ the determination of $w_0$ is 
uncorrelated with the other parameters, i.e.~the contours are vertical or 
horizontal. 

\section{Systematic Uncertainties and Observational Methods} \label{sec.sys}

Virtually all proposed observational methods for probing cosmology run into 
limits imposed 
by systematic uncertainties rather than statistical errors.  So although the 
cosmic shear test appears rather promising in its sensitivity to parameters and 
complementarity with other probes, as expected from its inclusion of a bare 
factor $H(z)$ as discussed in the first two sections, it behooves us not to 
make estimates of its power merely by speculating on achievable future survey 
statistics. 
For more realistic assessment of the promise of this method for determination of 
cosmological parameters, we must investigate the effect of systematics.  
Detailed 
characterization of irreducible uncertainities requires a comprehensive survey 
design and analysis; instead we present here a simple model that 
should illustrate the 
main effects and give reasonably realistic quantitative results. 

We adopt a precision of 2\% in measurement, with $n=10$ observations per 0.1 
redshift bin, but also include an irreducible systematics floor of 
2\% in a bin.  This makes the actual numbers used for the statistical error 
moot, except when we later consider a 
systematic that declines at low redshifts.  Generally we adopt a gaussian prior 
on the matter density of 0.03, but also investigated 0.01. The plots show 
results for observations over redshift ranges, e.g.~$z=1-2$. 

\begin{figure}[!hbt]
\begin{center} 
\psfig{file=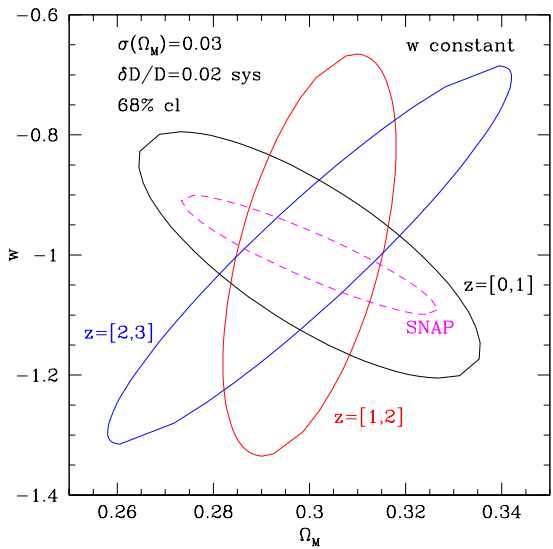,width=3in} 
\caption{Parameter estimation in the $w-\om$ plane, with systematics.  Note 
the equation of state is assumed constant a priori.  Contours correspond 
to the 68\% confidence level. 
} 
\label{fig.wom}
\end{center} 
\end{figure}

Figure \ref{fig.wom} shows the $w-\om$ plane, disallowing the possibility of any 
time variation in the equation of state -- an a priori assumption without 
justification. Note that systematics have a large effect on the \alp test. In 
particular they wipe out most of the 
complementarity at high redshift with the supernova method that was given by the 
$w$ sensitivity 
crossover and resulting rotation of the error ellipse.  (Note that the contours 
for the next generation supernova survey, 
SNAP, include systematics.) Some complementarity is retained for 
observations 
at $z>2$, allowing an improvement in determining $w$ and $\om$ by a factor two. 
Of course modeling the dark energy as a 
constant equation of state becomes even more suspect as one increases the range 
of 
observation.  

\begin{figure}[!hbt]
\begin{center} 
\psfig{file=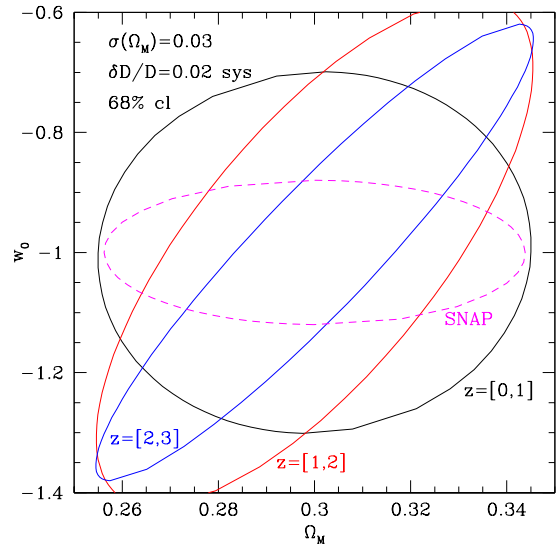,width=3in} 
\caption{Parameter estimation in the $w_0-\om$ plane, with systematics, 
marginalizing over $w'$.  Removing the prior that $w=$ constant strongly 
inflates the error contours.  
} 
\label{fig.w0om}
\end{center} 
\end{figure}

When one removes the constraint that the equation of state must be constant a 
priori, the \alp test does not offer any improvement to SNAP in estimation of 
the present value of the equation of state $w_0$, as seen in Figure 
\ref{fig.w0om}. 
The combination of the two experiments does help to limit $\om$, but so does, 
for example, the weak gravitational lensing survey that is an integral part of 
the SNAP mission. 

\begin{figure}[!hbt]
\begin{center} 
\psfig{file=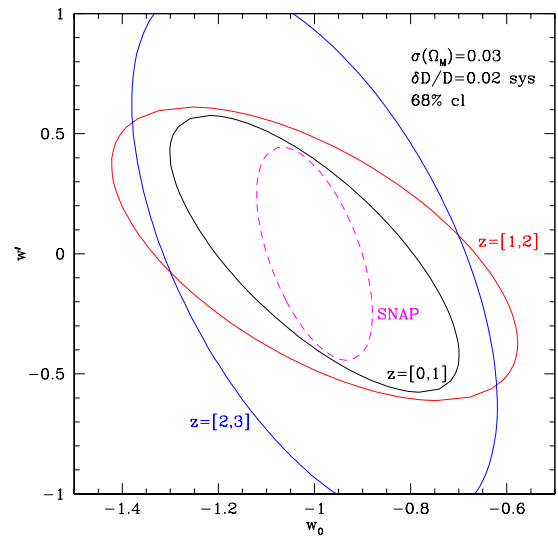,width=3in} 
\caption{Parameter estimation in the $w'-w_0$ plane, with systematics, 
marginalizing over $\om$.  Realistic assessment of systematics is key 
to evaluating the impact of the cosmic shear test.  
} 
\label{fig.wwp}
\end{center} 
\end{figure}

In the $w'-w$ plane (recall $w'\equiv w_a/2$), Figure \ref{fig.wwp} illustrates 
that 
not fixing $\om$ has a drastic effect on the error contours in comparison to the 
thin 
ellipses of Figure \ref{fig.idealwwp}. Uncertainty in $\om$ broadens the 
contours in 
a more or less fixed direction in this plane, though, so it affects the ellipses 
for some 
redshift ranges less than others. For example data around $z\approx0.7$ does not 
suffer as much loss of sensitivity to the equation of state parameters. Indeed 
observations covering the range of $z=0-1$ at 2\% precision permit 
determination of 
the time variation of the equation of state with errors about 30\% looser than 
from 
SNAP.  Unfortunately the 
respective error regions mostly overlap, with little complementarity.  Still, 
they 
would provide independent checks of the results using very different methods. 

In a bid to optimize the parameter determination we can examine how best to 
concentrate the observations, and over what redshift range. Given a systematics 
floor, 
an increase in the numbers of observations within a bin or an 
improvement in the statistical 
precision accomplishes little. An exception to this would be if external 
systematics 
needed to be reduced, for example by dividing the data into subsets for 
a ``like to 
like'' comparison. As for the redshift range, the results found here indicate 
that 
relatively little leverage is gained, for the given prior and systematics, 
outside 
the range $z\approx0.2-0.9$. 

This is well suited for ground based observations by large telescopes within a 
decade. 
Such a cosmic shear survey could be carried out by the KAOS project: the 
Kilo-Aperture Optical Spectrometer proposed as a front end for the 
8 meter Gemini South telescope. This would have multiplexing capability 
from some 4000  
apertures to measure detailed velocity maps of supercluster environments. In 
such a 
linear overdensity region one might hope to apply the \alp test without any 
complications of nonlinear gravitational physics or gas dynamics. Another 
possibility 
is studying the anisotropy of correlation functions of subclasses of bright 
galaxies. 
Both approaches can cover the preferred redshift range and statistics would not 
be a problem 
with a 1.5 square degree field of view and coverage of some 400 square degrees 
of sky 
to measure precise redshifts for $10^6$ galaxies. 

Observations at $z>1$ were shown to be fairly insensitive to the dark energy 
equation of state for 
the cosmic shear test, and thus can be used robustly to learn about 
astrophysics, perhaps from Lyman 
alpha forest observations.  Of course the rich panoply of data from 
such a next generation survey as KAOS can be examined with 
other cosmological probes as well (with similar cautions and care 
for the influence of systematic uncertainties).  

\section{Conclusion} \label{sec.concl}

The cosmic shear test looks extremely promising theoretically 
for the determination of cosmological parameters.  This is 
evident from its tomographic dependence on the expansion rate 
of the universe, shown by the appearance of the 
Hubble parameter $H(z)$ by itself.  It has further interesting properties 
in the redshift evolution of its parameter degeneracies and 
complementarity with other probes. 

However, systematic uncertainties, most probably involving 
peculiar velocities and distortions related to the local 
environment rather than cosmic expansion, put severe limits 
on the probe's ability to fulfill its potential. 

Unless systematic uncertainties can be brought under the 2\% level, 
the cosmic shear test applied at low or high redshift does not appear 
to offer significantly complementary or generally comparable 
limits to the supernova distance method.  Considering a linear 
systematic of $0.02z$, a matter density prior $\sigma(\Omega_m)=0.01$, 
or the addition of Planck CMB data  
does not greatly affect these conclusions. 

The best hope for applying this method does, 
however, lie at redshifts $z<1$, which is observationally feasible.  
Therefore careful 
study of the systematic uncertainties might yield some regime in which 
the next generation of wide, deep redshift surveys can bring this method 
of determining cosmological parameters to fruition. 

\section*{Acknowledgments} 

I thank Matthew Colless and Dragan Huterer for useful discussions. 
I acknowledge support for this work from the Director, Office of 
Science, DOE under DE-AC03-76SF00098 at LBL and by the NSF under 
PHY99-07949 at the KITP.

\end{document}